\documentclass[twocolumn,preprintnumbers,amsmath,amssymb]{revtex4}

\usepackage{graphicx}

\usepackage{dcolumn}

\usepackage{bm}

\begin{document}

\title{HIGHER DIMENSIONAL  BELL-SZEKERES METRIC }

\author{ M. G{\" u}rses} 

\altaffiliation
{email:
    gurses@fen.bilkent.edu.tr}

\affiliation{Department of Mathematics, Faculty of Sciences,
 Bilkent University, 06533 Ankara - Turkey}

\author{Y. \.Ipeko\u glu }

\altaffiliation{email: ipekoglu@metu.edu.tr}

\author{A. Karasu }

\altaffiliation
{email: karasu@metu.edu.tr}

\author{\c C. \c Sent{\"u}rk }

\altaffiliation
{email:  e126967@metu.edu.tr}

\affiliation{
 Department of Physics, Faculty of Arts and  Sciences,
Middle East Technical University, 06531 Ankara-Turkey}%


\begin{abstract}
The collision of pure electromagnetic plane waves with collinear polarization in $N$-dimensional 
$(N=2+n)$  Einstein-Maxwell theory is considered. A class of exact solutions
 for the higher dimensional Bell-Szekeres metric
is obtained  and its  singularity structure is examined.
\end{abstract}


\maketitle

\section{Introduction}

One of the main fields of interest in general relativity is the collision of gravitational plane waves.
Colliding plane wave space-times have been investigated in detail in general relativity \cite{gri}.
The first  exact solution of the Einstein-Maxwell equations representing
colliding plane shock electromagnetic waves with collinear polarizations was obtained by
Bell and Szekeres (BS) \cite{bel}. This solution is conformally flat in the interaction
region and its singularity structure has been considered by Matzner and Tipler 
 \cite{mat}, Clark and Hayward \cite{cla} and  Helliwell and Konkowski \cite{hel}. 
 Later Halil \cite{hal}, G{\" u}rses and Halil \cite{gur1},
Griffiths \cite{gri1} and Chandrasekhar and Xanthhopolos \cite{Chan}  studied
exact solutions of the Einstein-Maxwell equations describing the collision of gravitational
and electromagnetic waves. Furthermore G{\" u}rses and Sermutlu \cite{gur} 
, more recently Halil and
Sakalli \cite{hal1} have obtained the extensions of the BS solution
 in the Einstein-Maxwell-dilaton and Einstein-Maxwell-axion theories, respectively. 
Recently, we have given  a higher even dimensional extensions  of the vacuum colliding gravitational
waves with collinear and also combinations  with noncollinear polarizations. \cite{gur3,gur4}.

 In this work we give a higher dimensional formulation of BS metric. We present
an exact solution generalizing the BS  solution
and examine the singularity structure of the corresponding spacetimes
in the context of curvature and Maxwell invariants. We show that
this space-time, unlike BS metric, is not conformally flat.

In Sec. II,  we give a brief review of the BS solution and
in  Sec. III,  we formulate the $N$ dimensional
Einstein-Maxwell equations. In Sec. IV, we present the $N$ dimensional colliding exact
plane wave solutions describing the collision of shock electromagnetic waves. We also examine
the singularity structure of the corresponding space-times and show that interaction region
of our solution admits curvature singularities.

\section{The Bell-Szekeres Metric}
The BS metric is given by
\begin{equation}
ds^{2}=2dudv+e^{-U}(e^{V}dx^{2}+e^{-V}dy^{2})
\end{equation}
where the metric functions $U$ and $V$ depend on the null coordinates $u$ and $v$.
The electromagnetic vector potential has a single nonzero component $A=(0,0,0,A)$, where
$A$ is functions of $u$ and $v$.
The complete solution of the Einstein-Maxwell equations is

\begin{eqnarray}
&&U=-\log(f(u)+g(v)),~~A=\gamma (pw-rq), \nonumber\\
&&V=\log(rw-pq)-\log(rw+pq),
\end{eqnarray}
where 
\begin{eqnarray}
&&r=(\frac{1}{2}+f)^{1/2},~~~~ p=(\frac{1}{2}-f)^{1/2}, \nonumber\\
&&w=(\frac{1}{2}+g)^{1/2},~~~~ q=(\frac{1}{2}-g)^{1/2}
\end{eqnarray}
with
\begin{equation}
f=\frac{1}{2}-\sin^{2} P, ~~~~~~~g=\frac{1}{2}-\sin^{2} Q.
\end{equation}
Here $P=au\Theta(u)$,  $Q=bv\Theta(v)$, where  $\Theta$ 
is the Heaviside unit step function, $a$ and $b$ are arbitrary constants and
 $\gamma^{2}$=$\frac{8\pi}{\kappa}$
with $\kappa$ being the gravitational constant. The nature of the space-time singularity
 can be extracted from  the curvature invariant which  is a finite quantity
\begin{eqnarray}
&&I=R^{\mu\nu\alpha\beta}R_{\mu\nu\alpha\beta}\label{cu1} \\
&=&\frac{8}{w^{2}p^{2}q^{2}r^{2}(f+g)^{2}}[fg f_{u}^{2}g_{v}^{2}+ w^{2}p^{2}q^{2}r^{2}f_{uu}g_{vv}
\nonumber \\
&+&\frac{1}{4}((3g-f) r^{2}p^{2}f_{uu}g_{v}^{2}
+(3f-g) w^{2}q^{2}g_{vv}f_{u}^{2})]\nonumber\\
&=& 32 a^{2}b^{2}.\label{cu2}
\end{eqnarray}
Hence in the BS solution, the singularity that occurs when $f+g$=0 corresponds to a Cauchy
horizon. This solution is conformally flat in the interaction region; 
one of the components of the 
Weyl tensor 
\begin{equation}
C_{0202}=-\frac{wq}{2rp(rw+pq)^{2}(f+g)}[\frac{f}{r^{2}p^{2}}f_{u}^{2}+f_{uu}]
\end{equation}
shows that this solution possesses quasiregular singularities at the null boundaries \cite{cla}.
Finally, the invariant
\begin{equation}
F_{\alpha \beta}F^{\alpha \beta}=-2\gamma^{2} a b
\end{equation}
which is a finite quantity in the interaction region.
The BS solution in the interaction region is diffeomorphic to
the Bertotti -Robinson space-time \cite{gri,cla}.

\section{N- Dimensional Einstein-Maxwell Equations}

Let  $M$ be a $N=2+n$ dimensional manifold with a metric

\begin{eqnarray}
ds^2&=&g_{\alpha \beta}\,dx^{\alpha}\,dx^{\beta}\nonumber \\
&=&g_{ab}(x^{c})dx^a~dx^b+ H_{AB}(x^{c})dy^A~dy^B   \label{a0}
\end{eqnarray}

\noindent
where $x^{\alpha}=(x^{a},\, y^{A})$ , $x^{a}$ denote the local coordinates
on a 2-dimensional manifold and $y^{A}$ denote the local coordinates
on $n$-dimensional manifold, thus $a,b=1,2$, $A,B=1,2,...,n$.
The Christoffel symbols of the metric $g_{\alpha \beta}$ can be calcullated to give

\begin{equation}
\Gamma^{A}_{Ba}= \frac{1}{2}H^{AD}~~H_{DB,a},~~~
\Gamma^{a}_{AB}=- \frac{1}{2}g^{ab}~~H_{AB,b},
\end{equation}

\noindent

\begin{equation}
\Gamma^{A}_{BD}= \Gamma^{A}_{ab}=  \Gamma^{a}_{Ab}=0,\bar \Gamma^{a}_{bc} =\Gamma^{a}_{bc}
\end{equation}
\noindent
where  the $\Gamma^{a}_{bc}$ are the Christoffel symbols of the
2-dimensional metric $g_{ab}$.

\noindent
The components of the Riemann tensor are given by

\begin{equation}
R^{\alpha}_{ \beta \gamma \sigma}=\Gamma^{\alpha}_{\beta \sigma,  \gamma}-
\Gamma^{\alpha}_{\beta \gamma, \sigma}+\Gamma^{\alpha}_{\rho \gamma}\,
\Gamma^{\rho}_{\beta \sigma}-\Gamma^{\alpha}_{\rho \sigma}\,
\Gamma^{\rho}_{\beta \gamma}.
\end{equation}

\noindent
The components of the Ricci tensor are

\begin{eqnarray}
{\cal R}_{ab}&=&R^{\alpha}_{a \alpha b}
=R_{ab}+\frac{1}{4}tr(\partial_{a}H^{-1}\partial_{b}H)\\ \nonumber
&-&\bigtriangledown_{a} \bigtriangledown_{b}log \sqrt{detH},\\
{\cal R}_{AB}&=&- \frac{1}{2}(g^{ab}H_{AB,b})_{,a}
- \frac{1}{2}g^{ab}H_{AB,b}[\frac{(\sqrt detg)_{,a}}{ \sqrt detg}\nonumber\\
&+&\frac{(\sqrt detH)_{,a}}{ \sqrt detH}] 
+\frac{1}{ 2}g^{ab}H_{EA,b}H^{ED}H_{DB,a}, \\
{\cal R}_{aA}&= &0,
\end{eqnarray}

\noindent
where $R_{ab}$ is the Ricci tensor of the 2-dimensional metric $g_{ab}$.
The Maxwell potential 1-form ${\cal A}$ is 

\begin{equation}
{\cal A}=A_{A}dy^{A}.
\end{equation}

The components of the electromagnetic field

\begin{equation}
{\cal F}=\frac{1}{2}F_{\alpha \beta} dx^{\alpha}\wedge dx^{\beta}
\end{equation}
are
\begin{equation}
F_{aA}=A_{A,a},~~~~~~~~F_{ab}=0,~~~~~~~F_{AB}=0,
\end{equation}

The components of the energy-momentum tensor 

\begin{equation}
T_{\mu \nu}=\frac{1}{ 4 \pi}[g^{\alpha \beta} F_{\mu \beta}F_{\nu \alpha}-\frac{1}{4}g_{\mu \nu} F_{\alpha \beta}
F^{\alpha \beta}]
\end{equation}
are
\begin{eqnarray}
T_{a b}&=&\frac{1}{4\pi}[H^{A B} F_{a A}F_{b B}-\frac{1}{ 2}g_{a b} F^{2}],\nonumber\\
T_{AB}&=&\frac{1}{ 4\pi}[g^{ab} F_{a A}F_{bB}-\frac{1}{ 2}H_{AB} F^{2}],\nonumber\\
T_{aA}&=&0,
\end{eqnarray}
where $F^{2}=F_{aD}F^{aD}$.
Then the Einstein field equations are
\begin{equation}
R_{\mu \nu}=\kappa [T_{\mu \nu}+\frac{1}{ 2-N}g_{\mu \nu}T]
\end{equation}
where the trace of the energy momentum tensor $T$ is

\begin{equation}
T=\frac{1}{ 8\pi}(2-n)F^{2}.
\end{equation}

The Einstein-Maxwell equations are:

\begin{eqnarray}
&&R_{ab}+ \frac{1}{4}tr(\partial_{a}H^{-1}\partial_{b}H)-
\bigtriangledown_{a} \bigtriangledown_{b}\log \sqrt{\det{H}}\nonumber\\
&=&\frac{\kappa}{ 4 \pi}H^{AB} F_{aA} F_{bB}-
\frac{\kappa }{ 4 \pi n}g_{ab}F^{2} \label{E1},\\
&&\partial_{a}[\sqrt{\det{H}\,g}\;g^{ab}H^{AS}\partial_{b}H_{AB}]\nonumber \\
&=&-\frac{\kappa}{ 2 \pi} \sqrt{\det{H}\,g}\;[H^{AS}g^{ab}F_{Aa}
F_{Bb}-\frac{\delta^{S}_{B}}{ n}F^{2}]  \label{E2}
\end{eqnarray}
and
\begin{equation}
\partial_{a}[\sqrt{\det{H}\,g}\; F^{Aa}]=0, \label{E3}
\end{equation}
\noindent
where
 $\bigtriangledown$ is the covariant differentiation with respect
to the connection $\Gamma^{a}_{bc}$ (or with respect to metric
$g_{ab}$). We may rewrite the 2-dimensional metric as

\begin{equation}
g_{ab}=e^{-M}\,\eta_{ab} ,\;\;\;
\end{equation}

\noindent
where $\eta$ is the metric of flat 2-geometry with arbitrary signature
($0$ or $\pm 2$ ) the function $M$ depends on the local coordinates $x^{a}$.
The corresponding Ricci tensor and the Christoffel symbols  are

\begin{eqnarray}
R_{ab}&=& \frac{1}{ 2} (\bigtriangledown_{\eta}^{2} M)\,\eta_{ab},\nonumber \\
\Gamma^{c}_{ab}&=& \frac{1}{2}[-M_{,b}\delta^{c}_{a}
-M_{,a}\delta^{c}_{b}+M_{,d}\eta^{cd}\eta_{ab}].
\end{eqnarray}

\section{Higher Dimensional Bell-Szekeres metric}

In this section we give the higher dimensional colliding exact plane wave metric generalizing the
BS metric. For this 
purpose let $H$ be a diagonal matrix 
\begin{equation}
H=e^{-U}h
\end{equation}
where

$$h={\begin{pmatrix}e^{V_{1}} &&&\cr
&\ddots & &\bigcirc\cr
\bigcirc & &\ddots & \cr
& && e^{V_{n}} \cr \end{pmatrix}}$$

\noindent
with
 $ \det h=1$  i.e.,  $\sum_{k=1}^{n-1} V_{k}+V_{n}=0$.

Now taking the signature of flat-space metric with null coordinates
$$\eta=\left(\begin{array}{cc} 0&1\\1&0 \end{array}\;
\right),~~~ x^{1}=u,\, x^{2}=v,$$
and $$A_{A}=(0,...,A)$$  the Einstein-Maxwell equations become
\begin{eqnarray}
&&2U_{uv}-nU_{u}U_{v}=0, \label{aa1}  \\
&-&\frac{n}{2}U_{u}V_{kv}-\frac{n}{2}U_{v}V_{ku}+2V_{kuv}\nonumber\\
&=&\frac{\kappa}{\pi n}
e^{U-V_{n}}A_{u}A_{v}, \label{aa2} \\
&-&\frac{n}{2}U_{u}V_{nv}-\frac{n}{2}U_{v}V_{nu}+2V_{nuv}\nonumber\\
&&=\frac{\kappa (1-n)}{\pi n}
e^{U-V_{n}}A_{u}A_{v}, \label{aa3} \\
&&(\frac{n-2}{2})(U_{u}A_{v}+U_{v}A_{u})+V_{nu}A_{v}+V_{nv}A_{u}\nonumber\\
&&=2A_{uv},\\
&-&\frac{n}{2}U_{u}^{2}-\frac{1}{2}\sum_{k=1}^{n-1}(V_{ku})^{2}-\frac{1}{2}(V_{nu})^{2}
+nU_{uu}+nM_{u}U_{u}\nonumber\\
&=&\frac{\kappa}{2 \pi}e^{U-V_{n}}(A_{u})^{2}, \\
&-&\frac{n}{2}U_{v}^{2}-\frac{1}{2}\sum_{k=1}^{n-1}(V_{kv})^{2}-\frac{1}{2}(V_{nv})^{2}
+nU_{vv}+nM_{v}U_{v}\nonumber \\
&=&\frac{\kappa}{2 \pi}e^{U-V_{n}}(A_{v})^{2}, \\
&&2M_{uv}-\frac{n}{2}U_{u}U_{v}-\frac{1}{2}\sum_{k=1}^{n-1}V_{kv}V_{ku}-\frac{1}{2}V_{nv}V_{nu}
\nonumber\\
&+&nU_{uv}=\frac{\kappa}{2 \pi n}(2-n)e^{U-V_{n}}A_{u}A_{v},
\end{eqnarray}
where $k=1,...,n-1$. Note that the last equation  is not independent. It can be obtained 
from the other equations.
The most general solution to  Eq. (\ref{aa1}) is given by
\begin{equation}
U=-\frac{2}{n}\log [f(u)+g(v)]
\end{equation}
in terms of two arbitrary functions $f$ and $g$. Now changing variables $(u,v)$ to $(f,g)$ the
remaining field equations become
\begin{eqnarray}
&&-\frac{n}{2}U_{f}V_{kg}-\frac{n}{2}U_{g}V_{kf}+2V_{kfg}=\frac{\kappa}{\pi n}
e^{U-V_{n}}A_{f}A_{g},~~~~ \label{s1} \\
&-&\frac{n}{2}U_{f}V_{ng}-\frac{n}{2}U_{g}V_{nf}+2V_{nfg}=\frac{\kappa (1-n)}{\pi n}
e^{U-V_{n}}\nonumber \\
&&A_{f}A_{g}, \label{s2} \\
&&(\frac{n-2}{2})(U_{f}A_{g}+U_{g}A_{f})+V_{nf}A_{g}+V_{ng}A_{f}\nonumber\\
&=&2A_{fg},\label{s3}\\
&&M_{u}=-\frac{f_{uu}}{f_{u}}+\frac{(n-1)}{n}\frac{f_{u}}{f+g}\nonumber\\
&-&\frac{(f+g)}{4 f_{u}}[\sum_{k=1}^{n-1}(V_{ku})^{2}+(V_{nu})^{2}
+\frac{\kappa}{ \pi}e^{U-V_{n}}(A_{u})^{2}], \label{s4}\\
&&M_{v}=-\frac{g_{vv}}{g_{v}}+\frac{(n-1)}{n}\frac{g_{v}}{f+g}\nonumber\\
&-&\frac{(f+g)}{4 g_{v}}[\sum_{k=1}^{n-1}(V_{kv})^{2}+(V_{nv})^{2}
+\frac{\kappa}{ \pi}e^{U-V_{n}}(A_{v})^{2}] \label{s5}
\end{eqnarray}
\noindent
Eqs. (\ref{s1}), (\ref{s2}) and (\ref{s3}) are integrability conditions for
 Eqs. (\ref{s4}) and (\ref{s5}).
An exact solution to the above Eqs. (\ref{s1}), (\ref{s2}), and (\ref{s3}) is
\begin{eqnarray}
&&V_{k}=\alpha_{k}\log(rw-pq)+\beta_{k}
\log(rw+pq),\label{p3}\\
&&V_{n}=-\alpha \log(rw-pq)-\beta \log(rw+pq),\label{p4}\\
&&A=\gamma (pw-rq)
\end{eqnarray}
with
\begin{equation}
 \alpha_{k}-\beta_{k}=\frac{\kappa \gamma^{2}}{2 \pi n}\label{b1}
\end{equation}
for all $k=1,...,n-1$,  and
\begin{equation}
 \sum_{k=1}^{n-1}\alpha_{k}=\alpha=\frac{2}{n},~~~ 
 \sum_{k=1}^{n-1}\beta_{k}=\beta=\frac{2}{n}-2. 
\end{equation}
Then we may obtain the value of $\gamma$ as
$\gamma^{2}=\frac{4n\pi}{\kappa (n-1)}$.
 
It is convenient to put Eqs. (\ref{s4}) and (\ref{s5}) in the following form \cite{gri}
\begin{equation}
e^{-M}=\frac{f_{u}g_{v}}{(f+g)^{(n-1)/n}}e^{-S}
\end{equation}
where $ S $ satisfies

\begin{eqnarray}
&&S_{f}=-\frac{(f+g)}{4}
[\sum_{k=1}^{n-1}(V_{kf})^{2}+(V_{nf})^{2}\\ \nonumber
&+&\frac{\kappa}{\pi}e^{U-V_{n}}(A_{f})^{2}],\\
&&S_{g}=-\frac{(f+g)}{4}
[\sum_{k=1}^{n-1}(V_{kg})^{2}+(V_{ng})^{2}\\ \nonumber
&+&\frac{\kappa}{\pi}e^{U-V_{n}}
(A_{g})^{2}]
\end{eqnarray}

Therefore we may write the metric function $M$ as
\begin{widetext}
\begin{eqnarray}
&&M=-\log(cf_{u}g_{v})+[\frac{n-1}{n}-\frac{4+n^{2}m_{2}}{4n^{2}}]\log(f+g)
+(\frac{n}{4(n-1)}) \log(\frac{1}{2}-f) \nonumber\\
&&+(\frac{n}{4(n-1)})\log(\frac{1}{2}+f)
+   (\frac{n}{4(n-1)}) \log(\frac{1}{2}-g)
+ (\frac{n}{4(n-1)}) \log(\frac{1}{2}+g)\nonumber \\
&&+\frac{1}{8n}[8-4n+n(m_{1}-m_{2})]\log(1+4fg+4prwq)
\end{eqnarray}
\end{widetext}
where $c$ is a constant and
\begin{equation}
 \sum_{k=1}^{n-1}\alpha_{k}^{2}=m_{1},~~~ \sum_{k=1}^{n-1}\beta_{k}^{2}=m_{2}.
\end{equation}
 $m_{1}$ and $m_{2}$, using Eq. (\ref{b1}), satisfy
\begin{equation}
m_{1}+m_{2}-2m_{3}=\frac{4}{n-1},~~~m_{1}-m_{2}=\frac{4 (2-n)}{n(n-1)}
\end{equation}
with
 $$\sum_{k=1}^{n-1}\alpha_{k} \beta_{k}=m_{3}$$.

The metric function $e^{-M}$ must be continuos across the null boundaries. To make it so
we assume that the functions $f$ and $g$ take the form
\begin{equation}
f=\frac{1}{2}-\sin^{n_{1}}P,~~~~~~~~~~g=\frac{1}{2}-\sin^{n_{2}}Q.
\end{equation}
Then the metric function $e^{-M}$  is
continuous across the boundaries if
\begin{equation}
n_{1}=n_{2}=\frac{4(n-1)}{3n-4}.
\end{equation}
Therefore, the metric function $e^{-M}$ reads
\begin{widetext}
\begin{equation}
e^{-M}=\frac{(1+4fg+4pqrw)^{k_{1}}(1-(\frac{1}{2}-f)^{2/n_{1}})^{1/2}
 (1-(\frac{1}{2}-g)^{2/n_{1}})^{1/2}}{(\frac{1}{2}+f)^{1-\frac{1}{n_{1}}}
(\frac{1}{2}+g)^{1-\frac{1}{n_{1}}} (f+g)^{k_{2}}}\label{p2},
\end{equation}
\end{widetext}
where
\begin{equation}
k_{1}=\frac{n-2}{2(n-1)},~~~~~~~k_{2}=\frac{n-1}{n}-\frac{1}{4n^{2}}(4+n^{2}m_{1}).
\end{equation}
It may thus be observed that the constant $n_{1}$ (= $ n_{2}$)  is restricted to the range
satisfying
\begin{equation}
 2 \ge n_{1}=n_{2} >\frac{4}{3}.\label{p1}
 \end{equation} 
It is also appropriate to choose $c=\frac{1}{n_{1}^{2}}$.

The spacetime line element generalizing the BS metric in $N=2+n$ dimensions is
\begin{equation}
ds^{2}=2e^{-M}du~dv+e^{-U}(e^{V_{1}}dx_{1}^{2}+...+e^{V_{n}}dx_{n}^{2}),
\end{equation}
where the metric functions  are given in  Eqs. (\ref{p3}-\ref{p4}) and (\ref{p2}).
Because of  Eq. (\ref{p1}) the metric  we have found is $C^{1}$  for $n>2$
across the null boundaries. In spite of this fact, the Ricci tensor is regular across the null boundaries
due to the Einstein field equations.
The above solution reduces to the well known BS solution for $n=2$.

We now discuss the nature of the spacetime singularities.
We study the behavior of the metric functions $U$, $V_{k}$, $V_{n}$ and
 $M$ as $f+g$ tends to zero.
In the BS solution the collision of 
the two shock electromagnetic plane waves generates impulsive gravitational waves along
the null boundaries. It  is shown that, apart from the impulsive waves themselves,
by  virtue of Eq. (\ref{cu2})
the BS solution has no curvature singularities and the only singularites  are of the  quasiregular
type \cite{cla}. 
The curvature invariant  Eq. (\ref{cu1}) for $ n>2$ is 
\begin{equation}
I \sim e^{2M} \frac{(f_{u}g_{v})^{2}}{(f+g)^{4}}.
\end{equation}
as $f+g \rightarrow 0$.
Using $M$ from Eq. (\ref{p2}) we find
\begin{equation}
I \sim (f_{u}\,g_{v})^2\, (f+g)^{2k_{2}-4}
\end{equation}
as $f+g \rightarrow 0$. It is obvious that  spacetimes possess 
curvature singularities when $ k_{2}<2$ and their strength depend on  $n$ and $m_{1}$.

We also investigate the singularity structure of  spacetimes
in the context of the Maxwell invariants;
one of the  invariants is
\begin{eqnarray}
&&F_{\alpha \beta}F^{\alpha \beta}=-\frac{\gamma^{2}n_{1}^{2}}{2^{k_{1}+1}}(rwpq)^{1-\frac{2}{n_{1}}}
(rw+pq)^{-2k_{1}}\nonumber\\
&&\times(f+g)^{k_{2}}P_{u}Q_{v}.
\end{eqnarray}
which has singularities for  $n>2$ for the negative values of $k_{2}$.

We finally examine the Weyl tensor to see whether our spacetime is conformally flat.
One of
the components of the Weyl tensor in region II for our space-times is
\begin{eqnarray}
&&C_{0n0n}=\frac{f_{u}^{2}}{8(\frac{1}{2}+f)}
[-m_{1}+\frac{2}{n(n-1)}\nonumber\\
&+& \frac{2n}{(n-1)}(\alpha+\beta)^{3}
+\frac{(1-n)}{n}(\alpha+\beta)\nonumber\\
 &+&2(\alpha+\beta)\frac{(\frac{1}{2}-f)^{-1+2/n_{1}}(\frac{1}{2}+f)}{
(1-(\frac{1}{2}-f)^{2/n_{1}})}]\nonumber\\
&+&\frac{(\alpha+\beta)}{4}f_{uu}.
\end{eqnarray}
It can be seen that it vanishes only for $n=2$. Therefore, the
higher dimensional extensions of BS metric  are not conformally flat.

\section{conclusion}

In this paper, we give a higher dimensional generalization of BS metric which
describes the collision of pure electromagnetic plane waves with collinear polarization in all spacetime
dimensions.
The solution has two free parameters;  the spacetime dimension $N (=2+n)$
and  an arbitrary real number $m_{1}$.
We show that these spacetimes, unlike BS metric, are not conformally flat. We find that, even though
 purely electromagnetic plane wave
collision in four dimensional spacetime possesses no curvature singularities, in higher dimensions
there exist
 curvature singularities whose nature depend on the real number $m_{1}$ and the spacetime
dimension.

This work is partially supported by the Scientific and Technical
Research Council of Turkey (TUBITAK) and by Turkish Academy of Sciences (TUBA).

\end{document}